\documentstyle{lamuphys}
\makeatletter
\let\chapter\hid@chapter
\makeatother

\begin{document}
\pagenumbering{arabic}
\title{Testing models of radio source space density
evolution with the SUMSS survey}

\author{C. A.\,Jackson\inst{1} and J. V.\,Wall\inst{2}}

\institute{School of Physics, University of Sydney, NSW 2006
Australia
\and
Royal Greenwich Observatory, Madingley Road,
Cambridge CB3 0EZ, UK}

\titlerunning{radio source evolution \& SUMSS}
\maketitle

\begin{abstract}

We present the population data as predicted by
our space density analysis at the SUMSS frequency of 843 MHz.  
This data demonstrates the potential of the SUMSS survey to 
trace in detail the change-over in radio source populations
from `radio monsters' at high flux densities to a 
local starbursting population at the survey limit. The exact form
of the observed change-over will be used to refine our models
of radio source evolution. 
\end{abstract}

{\bf Extragalactic radio sources at 843 MHz} \\

 Our recent analyses (Wall \& Jackson 1997 and Jackson 1997) 
have shown that the powerful radio source
populations are well described by a dual-population unification
scheme. By adopting an elegantly simple aspect-dependent paradigm
we have determined that the
two `parent' populations have undergone quite separate
cosmic evolution histories with the result that their
space density distributions are very different.  
Whilst our working hypothesis is straightforward, the results from
our analysis are found to be comprehensively supported by 
cosmological tests embodied in radio source count and identification
data over a wide radio frequency range (151 MHz -- 8.4 GHz). 

The Sydney University Molonglo sky survey (`SUMSS', E. Sadler these
proceedings) at 843 MHz
can be considered to be an `intermediate'
radio frequency survey, lying between the `low' frequency surveys
($\nu <$ 200 MHz, {\it e.g.} 3C, 6C, 7C) and those at 
`high' frequency ($\nu >$ 2 GHz, {\it e.g.} Parkes, 87GB, PMN).  
The change in radio source types which comprise the source count between the
low and high frequencies has been shown to be a natural
consequence of the increasing contribution from the
preferentially-aligned sources which are Doppler-beamed along or very 
close to our line-of-sight.  
Surveys below $\sim$ 200 MHz are almost
completely dominated by extended radio galaxies whilst amongst the
brightest radio sources at 5 GHz
there are almost equal numbers of compact and extended sources. 

We have applied our dual-population 
space-density model to the SUMSS frequency
and show the predicted integral population mix in Figure 1.
The fraction of bright ($S_{843 {\rm \thinspace MHz}} >$ 1 Jy)
sources which will appear flat-spectrum 
($\alpha_{843 {\rm\thinspace MHz}}^{5 {\rm \thinspace GHz}} >$ -0.5)
is determined to be $\sim$ 10 \% from a fit to the 
well-defined source count at 1.4~GHz.
The changing contribution from each source type 
is a result of the differential evolution of the underlying
`parent' populations in the radio luminosity function. 

Figure 1 shows that at high flux density ($S_{843 
{\rm \thinspace MHz}} >$ 0.1 Jy) 
SUMSS will be dominated by the high-power population: FRII radio
galaxies with strong optical emission lines (spectral class `A', Hine
\& Longair 1979) and 
quasars, their beamed counterparts.  There is also a significant
contribution from the lower-power population:  
radio sources which have only weak,
if any, optical emission lines ({\it i.e.} 
FRIs and low-excitation FRIIs, spectral class `B', Hine \& Longair
1979 and Laing {\it et al.} 1994).
In contrast, at the milli-jansky flux density level the  
dominant populations are expected to be the
starburst and Seyfert galaxies, the low-excitation 
radio galaxies and their beamed counterparts  (BL Lac-type sources). 

\begin{figure}
\vspace{9.5cm}
\includegraphics{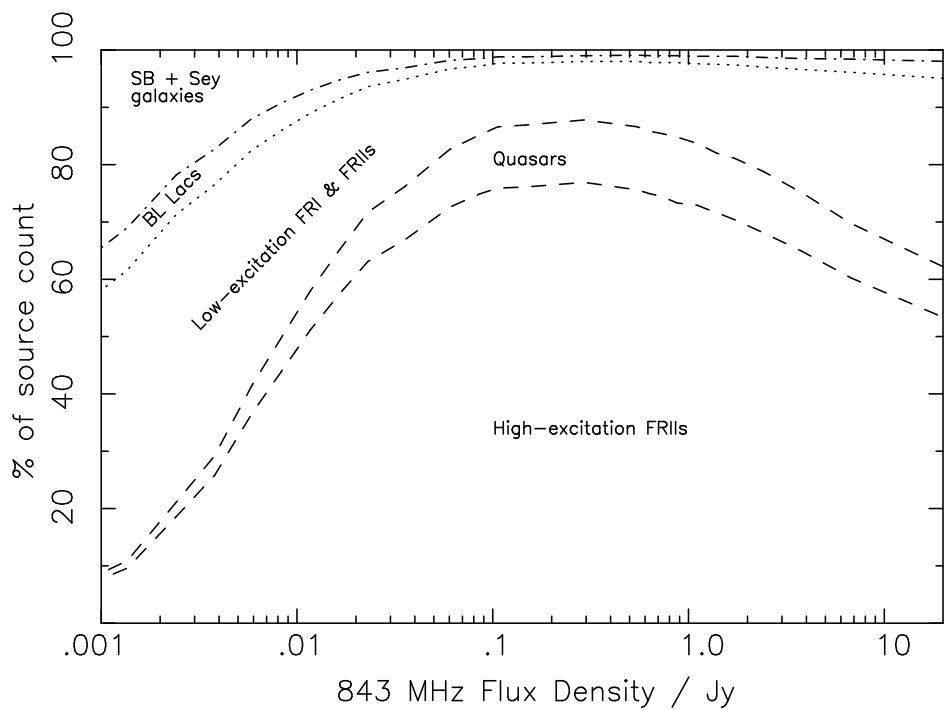}
\caption{Predicted population mix at 843 MHz as a function of flux density.}
\end{figure}

%
% ---- Bibliography ----
%


\begin{thebibliography}
%
\bibitem{}{hine1}{Hine 1979}
Hine, \, R. G., Longair, \, M. S. (1979)
%Optical spectra of 3CR radio galaxies.
MNRAS {\bf 188}, 111--130
%
\bibitem{}{jack1}{Jackson 1997}
Jackson, \, C. A. (1997)
%Cosmological tests of unified models for extragalactic 
%radio sources. 
PhD thesis, University of Cambridge
%
\bibitem{}{laing1}{Laing 1994}
Laing, \, R. A., Wall, \, J. V., Jenkins, \, C. R., 
Unger, \, S. W. (1994)
%Spectrophotometry of a complete sample of 3CR radio galaxies:
%implications for unified models.
{\it ASP Conf Ser Vol 54, The Physics of Active Galactic Nuclei}
(ASP, San Francisco), 201--208
%
\bibitem{}{wall1}{Wall \& Jackson 1997}
Wall, \, J. V., Jackson, \, C. A. (1997)
%Dual-population radio source unification.
MNRAS {\bf 290}, L17--L22
%

\end{thebibliography}
\end{document}